\documentclass[letterpaper]{article} 
\usepackage{aaai2026}  
\usepackage{times}  
\usepackage{helvet}  
\usepackage{courier}  
\usepackage[hyphens]{url}  
\usepackage{graphicx} 
\usepackage{natbib}  
\usepackage{caption} 
\usepackage{algorithm}
\usepackage{algorithmic}

\usepackage{booktabs}
\usepackage{multirow}
\usepackage{makecell}
\usepackage{amsmath}

\usepackage{newfloat}
\usepackage{listings}
\DeclareCaptionStyle{ruled}{labelfont=normalfont,labelsep=colon,strut=off} 
\floatstyle{ruled}
\newfloat{listing}{tb}{lst}{}
\floatname{listing}{Listing}
\title{AgentODRL: A Large Language Model-based Multi-agent System \\for ODRL Generation}
\author{
    Wanle Zhong,
    Keman Huang\protect\thanks{Corresponding author.},
    Xiaoyong Du
}
\affiliations {
    School of Information, Renmin University of China,
    Beijing, China\\
    \{wanle, keman, duyong\}@ruc.edu.cn
}

\nocopyright

\begin{document}

\maketitle

\begin{abstract}

The Open Digital Rights Language (ODRL) is a pivotal standard for automating data rights management. However, the inherent logical complexity of authorization policies, combined with the scarcity of high-quality ``Natural Language-to-ODRL" training datasets, impedes the ability of current methods to efficiently and accurately translate complex rules from natural language into the ODRL format. To address this challenge, this research leverages the potent comprehension and generation capabilities of Large Language Models (LLMs) to achieve both automation and high fidelity in this translation process. We introduce AgentODRL, a multi-agent system based on an Orchestrator-Workers architecture. The architecture consists of specialized Workers, including a Generator for ODRL policy creation, a Decomposer for breaking down complex use cases, and a Rewriter for simplifying nested logical relationships. The Orchestrator agent dynamically coordinates these Workers, assembling an optimal pathway based on the complexity of the input use case. Specifically, we enhance the ODRL Generator by incorporating a validator-based syntax strategy and a semantic reflection mechanism powered by a LoRA-finetuned model, significantly elevating the quality of the generated policies. Extensive experiments were conducted on a newly constructed dataset comprising 770 use cases of varying complexity, all situated within the context of data spaces. The results, evaluated using ODRL syntax and semantic scores, demonstrate that our proposed Orchestrator-Workers system, enhanced with these strategies, achieves superior performance on the ODRL generation task.

\end{abstract}

\begin{links}
    \link{Code}{https://github.com/RUC-MAS/AgentODRL}
\end{links}


\section{Introduction}

With the burgeoning development of the digital economy, the importance of Data Spaces as a distributed infrastructure to facilitate data exchange among multiple organizations, while ensuring the principles of data sovereignty and interoperability~\cite{otto2019designing}, has become increasingly prominent. In such ecosystems, to achieve trusted cross-entity data sharing and circulation~\cite{idsa_ram_4_0,farrell2023european}, initiatives like the International Data Spaces Association (IDSA) have adopted the W3C's Open Digital Rights Language (ODRL) as the core standard for describing usage policies for data assets~\cite{dam2023policy,meckler2023web,villata2018odrl,akaichi2024interoperable}. However, the creation of ODRL policies is deeply dependent on Semantic Web technologies, requiring policy authors to be familiar with the RDF graph data model, its serialization formats~\cite{kellogg2019json}, and the complex conceptual system of ODRL itself. This presents a significant barrier to adoption for domain experts without a technical background. Even today, with Large Language Models (LLMs) demonstrating powerful language understanding and generation capabilities, bridging this technical divide to accurately and reliably translate complex natural language (NL) rules from legal regulations and commercial agreements into structurally rigorous ODRL policies remains a core bottleneck.

Traditionally, methods for handling such translation tasks have tended to adopt a \textit{monolithic} architecture, relying on a single LLM to perform the entire end-to-end generation process from NL to ODRL. Such approaches, from the earlier ``Ontology-Guided" generation~\cite{kotis2020ontology} to subsequent methods that introduce ``Self-Correction Rules" (SCR) for optimization~\cite{mustafa2025instructions}, essentially attempt to address all types of complexity with a single, unified model. However, this architecture often proves inadequate when dealing with multiple, cognitively distinct sub-tasks concurrently—such as parsing dependencies in legal text, performing semantic segmentation, and executing constrained generation within a strict syntax. Its fundamental flaw lies in the inability of a single model to concurrently handle these diverse cognitive challenges. Consequently, its performance degrades significantly, especially when faced with policies containing complex logical structures like parallel or recursive dependencies, failing to guarantee the accuracy and completeness of the generated policies. This challenge is compounded by the scarcity of diverse training data \cite{han2023pive} and the dual need for syntactic and semantic fidelity.

To overcome the inherent limitations of the monolithic architecture, we note that the multi-agent system (MAS)~\cite{wang2024survey,hong2023metagpt,chen2024survey}, as a computational paradigm where multiple autonomous agents collaborate to solve complex problems, has shown great potential in numerous fields~\cite{hong2023metagpt,qian2023communicative,mandi2024roco,zhang2023building,xu2023language,park2023generative,boiko2023emergent}. Through task decomposition and collaboration, it can handle complex problems that are intractable for a single agent. However, the idea of applying a MAS to the specific task of automatic ODRL generation has not yet been fully explored. We posit that this architectural paradigm can offer a new perspective for resolving the complexity of ``NL-to-ODRL" conversion.

Therefore, this paper designs and implements AgentODRL, an integrated system that synergizes a multi-agent framework with targeted optimization strategies to address these challenges holistically. The system is based on an ``Orchestrator-Workers" pattern~\cite{dean2008mapreduce}. The core idea is a two-stage process: first, a central Orchestrator Agent analyzes the input's complexity and dispatches the task to specialized Worker Agents—a Rewriter for recursive structures and a Splitter for parallel ones. Second, after the Generator Agent produces an initial ODRL policy, two crucial post-generation strategies are employed to ensure quality: a validator-based iterative correction loop for syntactic accuracy and a LoRA-finetuned reflection mechanism for semantic fidelity.

The main contributions of this paper are as follows:

\begin{itemize}
    \item \textbf{We propose AgentODRL, an integrated framework combining a multi-agent architecture with dedicated syntactic and semantic enhancement strategies.} This design achieves state-of-the-art performance, improving the average grammar and semantic scores across all models by 5.39\% and 14.52\% respectively when compared to the SOTA strategy (SCR-Enhanced), all while ensuring near-perfect grammatical compliance.
    \item \textbf{We construct and release a new comprehensive benchmark dataset for supporting and evaluating ODRL generation}. We built a dataset of 770 use cases, each with an NL policy for conversion into ODRL. The dataset spans diverse structural complexities—including simple, parallel, and recursive forms—offering a robust foundation for evaluating our system and supporting future research in ODRL generation.
\end{itemize}


\section{Related Work}

\subsubsection{Automated Generation of ODRL Policies}

Research on converting NL policies into the ODRL format can be broadly divided into two phases. Early studies primarily relied on ontology extensions to manually or semi-automatically model rules from specific regulations, such as the General Data Protection Regulation (GDPR) or the European Union's Artificial Intelligence Act (EU AI Act)~\cite{de2019odrl,golpayegani2024aiup,esteves2021odrl}. This process requires significant manual intervention and is difficult to apply at scale.

The advent of LLMs introduced a new paradigm, with researchers exploring the use of pre-trained models like BERT for end-to-end knowledge graph construction~\cite{kumar2020building}. To enhance generation quality, subsequent studies have introduced methods such as SCR~\cite{mustafa2025instructions} or employed fine-tuned small models as validators~\cite{han2023pive,sean2023generating}. These explorations aim to optimize the performance of a single LLM in the end-to-end conversion task.

However, these methods still face core challenges to their practical application. First, high-quality ``NL-to-ODRL'' parallel corpora are extremely scarce. Related research confirms that data scarcity compels researchers to construct or augment datasets themselves~\cite{han2023pive}, which fundamentally limits the model's learning capabilities. Second, to circumvent the task's inherent difficulty, existing works generally confine their scope to policy texts with simple logical structures. They fail to systematically address real-world rules containing complex structures, such as parallel or recursive ones, hindering the applicability of their solutions in real-world scenarios.

\subsubsection{Multi-Agent Systems for Complex Task Decomposition}

To overcome the bottlenecks of a single LLM, LLM-based MAS have emerged as a new research paradigm. This paradigm solves problems intractable for a single model by decomposing complex tasks and enabling multiple agents, each playing a specific role, to collaborate and simulate collective intelligence~\cite{wang2024survey,hong2023metagpt,chen2024survey}. Currently, this ``divide-and-conquer" strategy has demonstrated great potential in diverse domains, including automated software development~\cite{hong2023metagpt,qian2023communicative}, multi-robot systems~\cite{mandi2024roco,zhang2023building}, scene simulation~\cite{xu2023language,park2023generative}, and scientific research~\cite{hong2023metagpt,boiko2023emergent}.

However, the application of this paradigm in domains requiring high precision and strict logic, such as data rights management~\cite{calvaresi2020personal}, remains in its nascent stages. The ODRL generation task, in particular, inherently integrates multiple cognitively distinct sub-tasks. This intrinsic complexity poses a fundamental challenge to traditional methods that rely on a single model. In contrast, the multi-agent architecture offers a highly aligned systemic solution through task decomposition and collaboration, revealing the significant potential of using architectural innovation to address the complexity of rule-based generation.


\section{Use Case Complexity in Practice}
\label{sec:use_case_classification}

To systematically process NL policies into machine-readable ODRL, we first establish a classification based on their structural complexity. Use cases are categorized into two main types according to their internal structural relationships: \textbf{Simple Use Cases} and \textbf{Complex Use Cases}. This classification depends on core semantic elements, referred to as ``rule information" ($\mathcal{R}_{info}$), which can be formally defined as a tuple:
\begin{equation*}
\mathcal{R}_{info} = (P, A, Ac, C_{policy}, C_{context})
\end{equation*}
where $P$ represents the \textbf{Party} (Assigner/Assignee); $A$ is the target \textbf{Asset}; $Ac$ is the permitted \textbf{Action}; $C_{policy}$ is the set of core policy clauses (\textbf{Permissions, Prohibitions, and Duties}); and $C_{context}$ is the set of contextual \textbf{Constraints}.

\subsection{Simple Use Cases}
\label{subsec:simple_cases}
Simple Use Cases are characterized as limited rule information with a monolithic structure. The essence of these use cases is the description of a self-contained policy or a few closely related policies, whose semantics and structure are highly independent without complex dependencies.

\begin{quote}
\textbf{Example 3.1: Data Access License}
\textit{``The DE\_Staatstheater\_Augsburg, a German cultural organization, manages the dataAPI `ShowTimesAPI'. This dataAPI holds valuable cultural assets. Policy regulates access to this dataAPI, granting subscribers like `Regional Newspaper', `Culture Research Institute', and `Cultural Platform Bavaria'. Access is restricted to Germany, and usage rights expire on May 10, 2025."}
\end{quote}

This example embodies a single, coherent policy. Given their self-contained nature and clear semantics, such simple use cases correspond to a baseline processing path in our proposed workflow. This path directly invokes the \textbf{Generator} agent (see Section \ref{sec:method}), typically without the need for complex structural preprocessing.

\subsection{Complex Use Cases}
\label{subsec:complex_cases}
In contrast to simple cases, the challenge of complex use cases lies not merely in their length but in the intricate structural relationships between their internal policies. We subdivide them into two subtypes: Use Cases with Parallel Structures and Use Cases with Recursive Structures.

\subsubsection{Use Cases with Parallel Structures}
\label{subsubsec:parallel}
Use Cases with Parallel Structures are characterized by the inclusion of multiple, relatively independent policies within a single use case. Such use cases are commonly found in standards, website terms of service, and comprehensive licensing frameworks. 

\begin{quote}
\textbf{Example 3.2: Tiered Access and Obligations}
\textit{``Registered non-commercial drone operators are permitted to download the `Alpine Geohazard Maps' from SwissTopo for flight planning, although access to the raw data files expires after 72 hours. While using the maps, they have an obligation to credit SwissTopo as the data source in their flight logs and are strictly prohibited from redistributing the original map files to third parties. For commercial companies seeking to integrate the data into proprietary software, access is granted upon payment of a per-square-kilometer fee."}
\end{quote}

Example 3.2 clearly contains two parallel policies for different user groups: an ``Offer" for non-commercial users and another for commercial companies. To faithfully convert such multifaceted use cases, a critical prerequisite is the logical decomposition of the source text into multiple, independent policy units. In our workflow, this task is undertaken by the \textbf{Splitter} agent (see Section \ref{sec:method}).

\subsubsection{Use Cases with Recursive Structures}
\label{subsubsec:recursive}
Use Cases with Recursive Structures are the most complex, featuring explicit cross-clause dependencies where one policy's enforcement is contingent on another. This structure is common in legal acts (e.g., GDPR, CCPA) and formal contracts.

\begin{quote}
\textbf{Example 3.3: Referential Dependency in Legal Clauses}
\textit{(From the Data Security Law of China)}

\textbf{Article 48:} \textit{``Whoever, in violation of the provisions of Article 35..., refuses to cooperate..., shall be fined..."}

\textbf{Article 35:} \textit{``When public security organs... retrieve data..., relevant organizations... shall cooperate."}
\end{quote}

The penalty in Article 48 is contingent on violating Article 35. This dependency, a hallmark of recursive structures, challenges automated conversion. It requires a rewriting stage by our \textbf{Rewriter} agent (see Section \ref{sec:method}) to resolve the reference, making the policy self-contained for subsequent processing.


\begin{figure*}[ht]
    \centering
    \includegraphics[width=\textwidth]{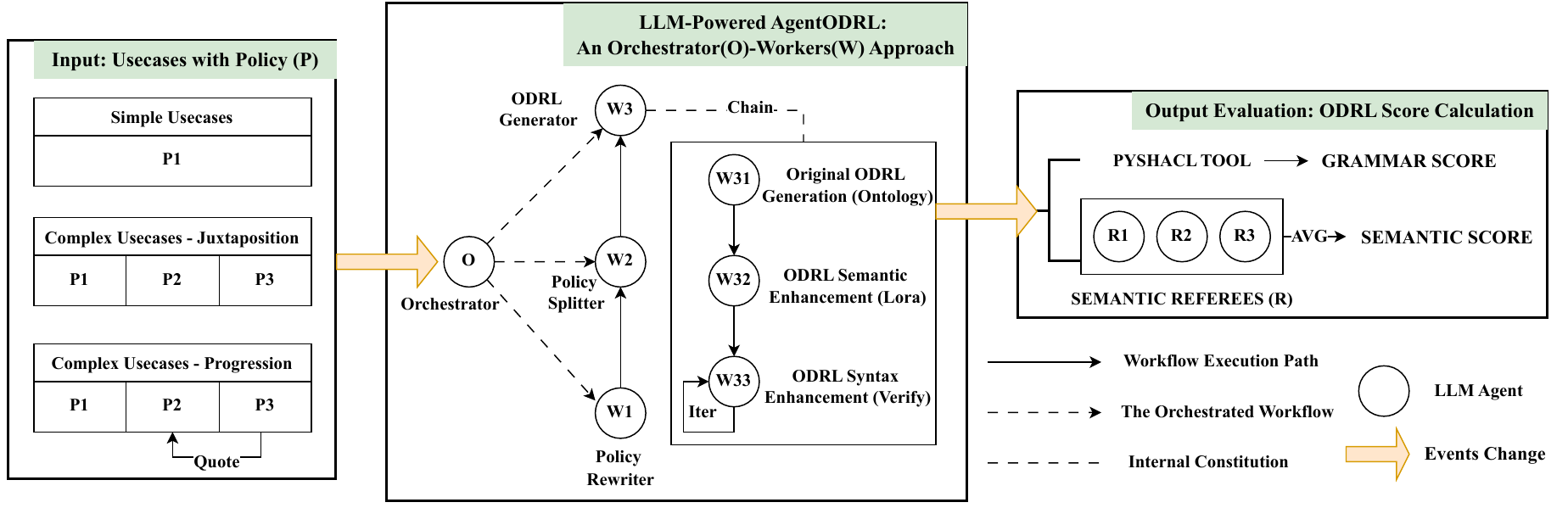}
    \caption{The ``Orchestrator-Workers" architecture of AgentODRL. The system takes various policy use cases as input (left panel), processes them through a central workflow orchestrated by an Orchestrator agent that delegates tasks to specialized Worker agents (center panel), and evaluates the final ODRL output using Grammar and Semantic scores (right panel).}
    \label{fig:agentodrl_arch}
\end{figure*}

\section{Methodology}
\label{sec:method}

This section details the design and implementation of \textbf{AgentODRL}, a MAS for converting NL data permission rules into the ODRL. The conversion is not a single, monolithic task. It inherently involves several distinct cognitive functions, including parsing complex logical structures, segmenting rules based on semantic intent, and generating code under strict syntactic constraints.

To effectively manage this complexity, AgentODRL employs a modular Orchestrator-Workers architecture. This design decouples the overall task, allowing specialized agents to handle specific cognitive functions where they excel. Such an approach enhances the workflow's robustness, interpretability, and maintainability, overcoming the limitations of a single-model approach. The architecture and working principle of this system are illustrated in Figure~\ref{fig:agentodrl_arch}.

\subsection{The Orchestrator-Workers Workflow}

AgentODRL adopts the ``Orchestrator-Workers” pattern, where a central LLM agent—the Orchestrator—acts as a project manager. It analyzes and decomposes tasks, assigning them to specialized ``worker” agents.

\textbf{Orchestrator Agent:} Serving as the central cognitive engine of the workflow, this agent receives NL use cases from the user. Its primary function is to rapidly analyze and classify the input based on the complexity criteria defined in the preceding chapter. Based on the classification result, the Orchestrator Agent determines which Worker Agents to invoke and orchestrates an optimal processing path.

\textbf{Worker Agents:} This component comprises a group of specialized agents, each engineered to handle a specific sub-task within the conversion process. Based on our analysis of ODRL policy complexity discussed in Section 3, this study implements three core Worker Agents: a \textbf{Rewriter Agent} for resolving cross-references in recursive structures, a \textbf{Splitter Agent} for decomposing parallel structures, and a \textbf{Generator Agent} for converting well-structured rule text into high-quality ODRL policies.

\subsection{Worker Agents in Detail}

\subsubsection{Rewriter Agent}

The Rewriter Agent addresses the most complex ``recursive structure" use cases by performing a ``Structure-Preserving Inlining" task. Guided by a structured prompt, it identifies and resolves both explicit (e.g., references to clause numbers) and implicit (e.g., ``Notwithstanding...") cross-references. The agent inlines the content of a referenced clause into the referring clause to eliminate semantic dependency. Crucially, this process preserves the structural separation of the original clauses, thereby simplifying the recursive problem into a set of independent statements prepared for the Splitter Agent.

\subsubsection{Splitter Agent}

The Splitter Agent is designed for ``parallel structure" use cases. Its core logic, encoded within a detailed prompt, directs its LLM to segment rules based on fundamental semantic shifts rather than superficial syntax. A new policy unit is initiated only upon a change in: (1) the core asset, (2) the core assigner-assignee relationship, or (3) the fundamental purpose of the policy. Following decomposition, the agent assigns an ODRL type (Agreement, Offer, or Set) via heuristics, ensuring each resulting unit is a self-contained and unambiguous input for the Generator Agent.

\subsubsection{Generator Agent}

The Generator Agent constitutes the core execution unit of the entire workflow. To guarantee the quality of the output, this agent integrates two key optimization strategies.

\paragraph{LoRA-Enhanced Strategy for Semantic Fidelity.}
To ensure the generated ODRL accurately reflects the intent of the original rule, we propose an innovative, reflection-based semantic optimization process. The core concept is a complementary ``generator-validator" model, which pairs a large generative model with a small, specialized ``expert validation model" fine-tuned with domain knowledge. We employ Low-Rank Adaptation (LoRA) to fine-tune a lightweight LLM (\texttt{Qwen3-4B-Instruct(BF16)}) into a high-precision, low-cost expert agent for semantic extraction. This expert agent extracts key semantic elements (e.g., parties, assets, actions) into a structured ``semantic checkpoint" list. The main Generator Agent must then validate its ODRL output against this list, ensuring every semantic point is accurately encoded.

In particular, the LoRA fine-tuning process utilized parameters of \texttt{$r$=16} and \texttt{$alpha$=32} on a dataset comprising 2,380 synthetic samples, the construction of which is consistent with the methodology used for creating use cases as detailed in the Section~\ref{Dataset_Construction}. Training was conducted on a single NVIDIA 4090 GPU for 3 epochs. The final model achieved a validation loss (\texttt{$eval\_loss$=0.0668}) that was significantly lower than the training loss (\texttt{$train\_loss$=0.129}), indicating effective generalization without overfitting and thus ensuring its reliability.

\paragraph{Validator-Based Strategy for Syntactic Correctness.}
To mitigate potential syntax errors when LLMs directly generate ODRL, we introduce a validator-based ``generate-validate-correct" iterative loop. Initially, the Generator Agent produces an ODRL policy. This policy is then submitted to a built-in validator, which is implemented using the PYSHACL library. This validator performs rigorous Shapes Constraint Language (SHACL)~\cite{knublauch2017shacl} validation against the ODRL information model and vocabulary. If validation fails, a detailed error report is fed back to the Generator Agent's LLM, prompting it to reflect and revise its output. This loop persists until the policy successfully passes validation or a preset maximum number of attempts is reached, thereby  enhancing the syntactic correctness of the final output.

\subsection{Workflow Paths}

For different use cases, AgentODRL’s adaptive workflow dynamically composes agents:

\begin{itemize}
    \item \textbf{Simple Path:} Input $\rightarrow$ Orchestrator $\rightarrow$ Generator $\rightarrow$ Output
    \item \textbf{Parallel Path:} Input $\rightarrow$ Orchestrator $\rightarrow$ Splitter $\rightarrow$ [Units] $\rightarrow$ Generator $\rightarrow$ [ODRLs] $\rightarrow$ Output
    \item \textbf{Recursive Path:} Input $\rightarrow$ Orchestrator $\rightarrow$ Rewriter $\rightarrow$ Splitter $\rightarrow$ [Units] $\rightarrow$ Generator $\rightarrow$ [ODRLs] $\rightarrow$ Output
\end{itemize}


\section{Dataset Construction}
\label{Dataset_Construction}
To the best of our knowledge, there exist no dataset for ODRL generation, mainly due to both the complexity of data right policy and ODRL. Given the lack of a suitable benchmark, we also constructed a new dataset tailored for data space scenarios to evaluate ODRL generation capabilities. The process involved two stages: Seed Case Formulation and LLM-based Data Augmentation.

\subsection{Stage 1: Seed Case Formulation}

We initially formulated 70 seed use cases, categorized by logical structure: simple (40), parallel (20), and sequential-dependency (10). The sources for these cases were diverse, including academic literature \cite{mustafa2025instructions}, drafts of industry standards\footnote{China Electronics Standardization Institute (CESI), \textit{General Requirements for Trusted Data Space Operation Rules}.}, and standard license agreements like Creative Commons 4.0. For the structurally complex sequential-dependency cases, we manually curated robust logical templates from legal texts such as the GDPR\footnote{Regulation (EU) 2016/679 of the European Parliament and of the Council of 27 April 2016 (\textit{GDPR}). \url{https://eur-lex.europa.eu/eli/reg/2016/679/oj}} and the California Consumer Privacy Act (CCPA)\footnote{California Consumer Privacy Act (CCPA) of 2018, Cal. Civ. Code §§ 1798.100 et seq. \url{https://leginfo.legislature.ca.gov/faces/billTextClient.xhtml?bill_id=201720180AB375}}.

\subsection{Stage 2: LLM-based Data Augmentation}
We then employed Gemini 2.5 Pro to augment the 70 seed cases, guided by two strict principles: \textbf{Core Logic Preservation} and \textbf{Contextual Element Transformation}. The former maintained the fundamental policy structure (rule type, action-target binding, constraint categories), while the latter diversified contextual elements (participants, data assets, use cases, etc.) to simulate real-world business variety.

Following this strategy and a subsequent comprehensive manual review, we constructed a high-quality dataset of 770 test cases, comprising \textbf{440 simple cases}, \textbf{220 parallel-relationship cases}, and \textbf{110 sequential-dependency cases}. This dataset establishes a robust foundation for evaluating our model and facilitating future research.


\section{Experiments}

This section presents two experiments conducted on our constructed dataset. The results demonstrate that our proposed ODRL generation strategy is state-of-the-art on both evaluation metrics, and that the modules of AgentODRL are highly efficient and effective in handling rule conversion.

\begin{table*}[t]
\centering
\small
\setlength{\tabcolsep}{2.5pt} 
\begin{tabular}{@{}cc cccc cccc c@{}}
\toprule
& & \multicolumn{4}{c}{\textbf{Grammar score}} & \multicolumn{4}{c}{\textbf{Semantic score}} & \multicolumn{1}{c}{\textbf{\makecell{Reflections(Avg)}}} \\
\cmidrule(lr){3-6} \cmidrule(lr){7-10}
\multicolumn{1}{c}{\textbf{Model}} & \multicolumn{1}{c}{\textbf{Use Cases}} & \textbf{OGS} & \textbf{\makecell{SCR-\\Enhanced}} & \textbf{\makecell{Semantic-\\Enhanced}} & \textbf{AOFP} & \textbf{OGS} & \textbf{\makecell{SCR-\\Enhanced}} & \textbf{\makecell{Semantic-\\Enhanced}} & \textbf{AOFP} & \textbf{\makecell{(For AOFP)}} \\ \midrule
\multirow{4}{*}{\textbf{GPT-4.1}}
& ALL Use Cases & 82.07 & 93.08 & \textit{\underline{93.47}} & \textbf{99.89} & 89.59 & 92.00 & \textit{\underline{97.78}} & \textbf{97.93} & 1.29 \\
& Simple Use Cases & 81.52 & 92.70 & \textit{\underline{93.59}} & \textbf{99.94} & 94.49 & 95.99 & \textbf{98.78} & \textit{\underline{98.64}} & 1.22 \\
& Complex - Parallel & 81.13 & 92.23 & \textit{\underline{92.52}} & \textbf{99.75} & 86.47 & 90.53 & \textit{\underline{96.47}} & \textbf{97.27} & 1.62 \\
& Complex - Recursive & 86.18 & \textit{\underline{96.32}} & 94.91 & \textbf{99.86} & 76.18 & \textit{\underline{78.97}} & \textbf{96.40} & \textbf{96.40} & 0.89 \\ \midrule
\multirow{4}{*}{\textbf{GPT-4.1-mini}}
& ALL Use Cases & 82.41 & 94.83 & \textit{\underline{95.44}} & \textbf{99.36} & 86.53 & 88.93 & \textit{\underline{96.25}} & \textbf{96.43} & 2.12 \\
& Simple Use Cases & 83.65 & 96.14 & \textit{\underline{96.99}} & \textbf{99.55} & 89.79 & 91.86 & \textit{\underline{97.50}} & \textbf{97.80} & 1.44 \\
& Complex - Parallel & 79.73 & 91.85 & \textit{\underline{92.65}} & \textbf{98.98} & 87.68 & 90.79 & \textit{\underline{94.74}} & \textbf{94.97} & 3.49 \\
& Complex - Recursive & 82.81 & \textit{\underline{95.94}} & 94.81 & \textbf{99.36} & 71.20 & 73.47 & \textbf{94.26} & \textit{\underline{93.85}} & 2.08 \\ \midrule
\multirow{4}{*}{\textbf{GPT-4.1-nano}}
& ALL Use Cases & 79.77 & \textit{\underline{88.40}} & 87.49 & \textbf{92.01} & 50.51 & 56.23 & \textit{\underline{65.67}} & \textbf{72.35} & 6.64 \\
& Simple Use Cases & 80.02 & \textit{\underline{89.32}} & 87.80 & \textbf{92.13} & 55.17 & 61.03 & \textit{\underline{70.38}} & \textbf{76.50} & 6.63 \\
& Complex - Parallel & 79.18 & \textit{\underline{87.25}} & 86.55 & \textbf{92.26} & 49.00 & 55.34 & \textit{\underline{61.00}} & \textbf{69.44} & 6.29 \\
& Complex - Recursive & 79.98 & \textit{\underline{89.02}} & 88.10 & \textbf{91.07} & 34.87 & 40.40 & \textit{\underline{56.20}} & \textbf{61.53} & 7.32 \\
\bottomrule
\end{tabular}
\caption{Complete results for Experiment 1, evaluating ODRL generation strategies across different GPT models and use case complexities. Our AgentODRL's Full Pipeline consistently outperforms both the Ontology-Guide Strategy and SCR-Enhanced methods across all models and complexity levels, with particularly strong gains in semantic accuracy for complex cases and near-perfect grammar scores. For each use case, the best-performing grammar and semantic scores are shown in bold, while the second-best scores are presented in \textit{\underline{italic and underlined}} format.}
\label{tab:exp1_results}
\end{table*}

\begin{table*}[t]
    \centering
    \small
    \setlength{\tabcolsep}{4pt}
    \begin{tabular}{@{}c c ccc c@{}}
        \toprule
        \textbf{Use Cases} & \textbf{Workflows} & \textbf{\makecell{Grammar Score}} & \textbf{\makecell{Semantic Score}} & \textbf{\makecell{Reflections(Avg)}} & \textbf{Tokens} \\
        \midrule

        \multirow{4}{*}{\textbf{Simple Use Cases}}
        & ODRL Generator & 92.13 & 76.50 & 6.63 & - \\
        & Splitter $\rightarrow$ Generator & \textbf{94.03} & \textbf{85.18} & \textbf{5.54} & - \\
        & Rewriter $\rightarrow$ Splitter $\rightarrow$ Generator & 93.56 & 84.50 & 5.54 & - \\
        & Orchestrator - Workers & 93.45 & 82.12 & 5.82 & - \\
        \midrule

        \multirow{4}{*}{\textbf{Complex - Parallel Structures}}
        & ODRL Generator & 92.26 & 69.44 & 6.29 & - \\
        & Splitter $\rightarrow$ Generator & \textbf{93.77} & \textbf{84.88} & \textbf{6.13} & - \\
        & Rewriter $\rightarrow$ Splitter $\rightarrow$ Generator & 92.66 & 83.30 & 6.44 & - \\
        & Orchestrator - Workers & 91.11 & 82.47 & 6.34 & - \\
        \midrule

        \multirow{4}{*}{\textbf{Complex - Recursive Structures}}
        & ODRL Generator & 91.07 & 61.53 & 7.32 & - \\
        & Splitter $\rightarrow$ Generator & 91.62 & 77.70 & 6.61 & - \\
        & Rewriter $\rightarrow$ Splitter $\rightarrow$ Generator & \textbf{93.27} & \textbf{82.00} & \textbf{6.03} & - \\
        & Orchestrator - Workers & 89.45 & 80.97 & 6.33 & - \\
        \midrule

        \multirow{4}{*}{\textbf{ALL Use Cases}}
        & ODRL Generator & 92.01 & 72.35 & 6.64 & 33856708 \\
        & Splitter $\rightarrow$ Generator & 93.62 & 84.02 & 5.86 & 47917323 \\
        & Rewriter $\rightarrow$ Splitter $\rightarrow$ Generator & 93.27 & 88.07 & 6.09 & 49451650 \\
        & Orchestrator - Workers & 92.56 & 80.22 & 6.04 & 46209570 \\
        \bottomrule
    \end{tabular}%
    \caption{Performance comparison of different workflow paths on the GPT-4.1-nano model. The table details results across specific use case categories and provides an overall summary for all use cases, including token consumption. The results demonstrate the necessity of aligning workflow paths with use case complexity. The best-performing path for each use case category is highlighted in bold.}
    \label{tab:workflow_comparison_merged}
\end{table*}

\subsection{Evaluation Metrics and Calculation Methods}

To comprehensively and quantitatively evaluate the performance of the proposed AgentODRL system and its various strategies on the NL-to-ODRL conversion task, we designed two core evaluation metrics: the \textbf{Grammar Score} and the \textbf{Semantic Score}.

\subsubsection{Grammar Score}

The Grammar Score measures whether a generated ODRL policy strictly adheres to the specifications of the W3C ODRL Information Model and Vocabulary. Syntactic correctness is a fundamental prerequisite for a policy's validity. To calculate this score, we employ an automated workflow based on SHACL. The process begins by establishing a comprehensive set of SHACL constraint rules based on the official ODRL standard. Subsequently, each generated ODRL policy is validated against this predefined ruleset to count the number of constraint violations ($N_{\text{errors}}$). The final Grammar Score is computed using the following formula:
\begin{equation*}
    \text{Score}_{\text{grammar}} (\%) = \left( 1 - \frac{N_{\text{errors}}}{N_{\text{constraints}}} \right) \times 100\%
\end{equation*}
where $N_{\text{constraints}}$ is the total number of constraints in the SHACL ruleset, and $N_{\text{errors}}$ is the total number of errors reported by the validator.

\subsubsection{Semantic Score}

The Semantic Score quantifies how faithfully a generated ODRL policy reflects the complete intent of the original rule. The assessment of this metric hinges on establishing a ``semantic checkpoint list", which comprises all indivisible, independent semantic points from the original rule.

The metric is assessed through an automated workflow based on an LLM Jury. This workflow first involves a dedicated ``Identifier" LLM to extract a ground-truth ``semantic checkpoint list" from the original NL use case. Subsequently, to ensure objectivity, a ``Jury" composed of two independent LLMs ($K=2$) meticulously compares the generated ODRL policy against the checkpoint list. Each juror assesses how accurately each semantic unit is reflected and assigns a score ($S_{ij}$). The final Semantic Score is calculated as the average of scores given by all jurors across all semantic units, according to the formula:
\begin{equation*}
\begin{split}
    \text{Score}_{\text{semantic}} (\%) & = \frac{1}{N_{\text{units}}} \sum_{i=1}^{N_{\text{units}}} \left( \frac{1}{K} \sum_{j=1}^{K} S_{ij} \right) \times 100\%
\end{split}
\end{equation*}
where $N_{\text{units}}$ is the total number of semantic units, $K$ is the number of jurors, and $S_{ij}$ is the score given by the $j$-th juror for the $i$-th semantic unit. 

\subsection{Experiment 1: Evaluating ODRL Generation Strategies}

To systematically evaluate the effectiveness of our proposed strategies, we consider the following two baselines:
\begin{itemize}
    \item \textbf{Ontology-Guided Strategy (OGS)}: This approach utilizes the ODRL v2.2 ontology \cite{odrl_ontology_v2_2} to guide LLMs in generating the ODRL policy. 
    \item \textbf{SCR-Enhanced} \cite{mustafa2025instructions}: Building on OGS, this approach uses predefined rules, sourced from the ODRL W3C recommendation and ontology relationships, to fix errors in the LLM's output and ensure it better aligns with the official specification.
\end{itemize}

Furthermore, we evaluate our framework progressively by considering two strategies based on whether they incorporate our proposed semantic and syntactic enhancements:
\begin{itemize}
    \item \textbf{Semantic-Enhanced Strategy}: This strategy integrates our LoRA-finetuned model to enhance semantic fidelity.
    \item \textbf{AgentODRL's Full Pipeline (AOFP)}: This represents our full, integrated system. It combines the Semantic-Enhanced Strategy with a validator-based strategy for syntactic correctness.
\end{itemize}

All tests were conducted on the GPT-4.1 series models (GPT-4.1, GPT-4.1-mini, and GPT-4.1-nano). Based on preliminary experiments, the maximum number of reflections was set to 8, a configuration applied to all experiments in this paper. To ensure the robustness of the results and mitigate the effects of randomness, each experimental configuration was run three times, with the final results averaged. 

As reported in Table~\ref{tab:exp1_results}, \textbf{our AOFP significantly outperforms the existing SCR-Enhanced strategy across all tested dimensions}. Its superiority is manifested in two key areas. First, our strategies significantly enhance both grammar and semantic fidelity. On the ``ALL Use Cases'' setting, our AOFP framework improves the average grammar score by 5.39\% and the average semantic score by a notable 14.52\% across the three models when compared to the SCR-Enhanced strategy. The peak improvement is even more pronounced, with the grammar score increasing by up to 7.32\% (for GPT-4.1) and the semantic score by up to 28.67\% (for GPT-4.1-nano). Second, the syntax strategy effectively ensures ODRL compliance. The ``generate-validate-reflect" closed-loop correction mechanism ensures the final output strictly adheres to W3C specifications. Within the AOFP, the grammar scores for all models rose to near-perfect levels (mostly exceeding 99), effectively rectifying syntactic hallucinations from the LLM.

Additionally, \textbf{our proposed strategies exhibit a generalizable enhancement effect across models of varying capabilities}. While a model's foundational capabilities determine the performance ceiling (GPT-4.1 $>$ GPT-4.1-mini $>$ GPT-4.1-nano), our strategies notably raise the performance floor of smaller models. For example, when using the Baseline strategy, GPT-4.1-nano scored a mere 34.87 in semantics on recursive structure use cases. However, our AOFP elevates this score to 61.53, achieve a 76.46\% improvement.

Moreover, \textbf{our strategy exhibits remarkable stability against increasing use case complexity}. As complexity escalates from ``Simple" to ``Recursive" cases, the performance of the OGS degrades sharply, whereas our AOFP maintains a high degree of stability (e.g., the semantic score for GPT-4.1 only slightly decreased from 98.64 to 96.4). This ability to maintain high performance is also reflected in the strategy's adaptive correction cost. The ``Average number of reflections" metric reveals the iterative effort expended by the syntax strategy to achieve compliance. The data show that weaker models and more complex use cases require more reflections (e.g., GPT-4.1-nano averaged 7.32 reflections for recursive cases). This capacity to sustain high performance under pressure while adaptively adjusting its process ensures the strategy's reliability in handling the diverse and complex rules found in real-world scenarios.

\subsection{Experiment 2: Orchestrator-Workers Workflow Performance Evaluation}

To comprehensively evaluate the overall efficacy of our proposed adaptive Orchestrator-Workers workflow, we designed Experiment 2. This experiment aims to validate its core value in a two-tiered process: first, by forcing all use cases through different fixed workflow paths, we establish the performance ``theoretical ceiling" for each complexity level as a benchmark. Second, we test whether the fully automated Orchestrator-Workers workflow can, without human intervention, approach this theoretical ceiling through dynamic analysis and intelligent routing. 

To more clearly observe the performance gains, we selected the GPT-4.1-nano model, which exhibited greater potential for improvement, as the execution agent for the Generator module. Concurrently, to maximize the analytical capabilities of the upstream modules, the Orchestrator, Splitter, and Rewriter agents all utilized the more powerful GPT-4.1 model. To ensure the stability and reproducibility of the results, each experimental configuration was run three times, with the final results averaged.

\subsubsection{Establishing the Performance Benchmark: The Role of Specialized Agents}
The data in Table~2 clearly reveals that the optimal processing path is dictated by the task's structural complexity, highlighting the distinct roles of our specialized worker agents. 

First, \textbf{the Splitter agent, designed for decomposing parallel policy structures, demonstrates universal benefits}. Across all use case categories, the workflow path that incorporates the Splitter before the Generator yields significant performance gains over the ``ODRL Generator". For instance, in ``Complex - Parallel Structures", the Semantic Score dramatically increases from 69.44 to 84.88. 

Second, \textbf{the Rewriter agent proves its value in more complex scenarios}. While the full workflow path, which includes the Rewriter, Splitter, and Generator in sequence, enhances performance across the board, its indispensable role is most evident in recursive use cases. Here, it achieves the highest possible Semantic Score of 82.00, effectively resolving the cross-clause dependencies that the baseline model struggles with. These findings validate a core principle: \textbf{performance optimization stems from the precise alignment of the processing path with task complexity}.

\subsubsection{Orchestrator's Intelligent Decision-Making: Balancing Performance and Efficiency}
The established benchmarks provide a clear frame of reference for evaluating the Orchestrator's value in automating workflow selection. As shown in Table~2, the Orchestrator-Workers workflow achieves an average Semantic Score of 80.22 across all use cases. This performance is not only significantly better than the ODRL Generator's score of 72.35 but also \textbf{demonstrates the effectiveness of the multi-agent collaborative architecture}.

While the Orchestrator's average score is slightly below the theoretical ceilings set by manually-selected, fixed workflows (e.g., 84.02 for the path using the Splitter, and 88.07 for the path using the Rewriter), its primary advantage lies in its efficiency. The Orchestrator intelligently routes tasks, thereby optimizing computational resource usage. The data on token consumption confirms this: the Orchestrator-Workers workflow required 46.2M tokens, which is more economical than forcing all use cases through either the fixed path utilizing the Splitter (47.9M tokens) or the full path initiated by the Rewriter (49.5M tokens). \textbf{This demonstrates the Orchestrator's success in automating the ``optimal choice" process—dynamically balancing near-optimal performance with significantly lower computational cost, making it a practical and potent solution for real-world applications}.


\section{Conclusion}

This paper introduced AgentODRL, a novel multi-agent system for converting complex NL policies into the machine-readable ODRL format. Experiments on our newly constructed 770-case benchmark demonstrate that: 1) our integrated generation strategies achieve state-of-the-art grammatical and semantic scores, and 2) the ``Orchestrator-Workers" architecture is highly effective at automating the selection of optimal processing paths. These findings establish AgentODRL as a potent solution for the challenges of real-world data policy automation.

\section*{Acknowledgments}

The work was supported by the National Natural Science Foundation of China (62441230, 62172425), and the Fundamental Research Funds for the Central Universities and the Research Funds of Renmin University of China (22XNKJ04).

\end{document}